\documentclass{emulateapj}
\slugcomment{Appeared in the Astrophysical Journal }

\newcommand{\kms}{\mbox{km\thinspace s$^{-1}$}}     

\newcommand{\msun}{\mbox{$M_\odot$}}

\shortauthors{Murphy et al.}
\shorttitle{Fokker-Planck Models for M15}

\begin{document}

\title{Fokker-Planck Models for M15 without a Central Black Hole: The
  Role of the Mass Function}

\author{Brian W. Murphy} \affil{Department of Physics and Astronomy,
  Butler University, Indianapolis, IN 46208}
\email{email: bmurphy@butler.edu}

\and

\author{Haldan N. Cohn and Phyllis M. Lugger}
\affil{Department of Astronomy, Indiana University, Bloomington, IN
  47405}

\begin{abstract}
We have  developed a set of dynamically  evolving Fokker-Planck models
for  the  collapsed-core globular  star  cluster  M15, which  directly
address the issue  of whether a central black hole  is required to fit
\emph{Hubble  Space  Telescope  (HST)}  observations  of  the  stellar
spatial distribution and kinematics.  As in our previous work reported
by  Dull et al.,  we find  that a  central black  hole is  not needed.
Using local  mass-function data from \emph{HST} studies,  we have also
inferred the  global initial stellar mass function.   As a consequence
of extreme mass segregation, the local mass functions differs from the
global mass  function at every  location.  In addition  to reproducing
the observed mass functions, the  models also provide good fits to the
star-count  and velocity-dispersion profiles,  and to  the millisecond
pulsar  accelerations.  We  address concerns  about the  large neutron
star  populations adopted  in  our previous  Fokker-Planck models  for
M15\@.  We  find that good model fits  can be obtained with  as few as
1600 neutron stars; this corresponds to a retention fraction of 5\% of
the initial  population for our  best fit initial mass  function.  The
models contain a substantial  population of massive white dwarfs, that
range in  mass up  to $1.2~\msun$.  The  combined contribution  by the
massive  white dwarfs  and  neutron stars  provides the  gravitational
potential needed  to reproduce \emph{HST} measurements  of the central
velocity dispersion profile.

\end{abstract}  

\section{Introduction}

\subsection{Motivation}
As  recently  reviewed by  \citet{vesperini10},  the possibility  that
globular  clusters may contain  intermediate-mass black  holes (IMBHs)
has received considerable attention,  particularly in the past decade.
An extrapolation of  the relation between central black  hole mass and
velocity  dispersion ($M_{\rm BH}-\sigma$)  for galaxies  implies that
globular clusters may  harbor black holes with masses  in the range of
$10^2-10^4\,\msun$.  There have been  a number of suggestions that the
prototypical collapsed-core globular cluster M15 may contain a central
black hole with mass of  near $10^3\,\msun$.  A central black hole has
been  invoked in M15  to explain  both the  central surface-brightness
cusp \citep{newell76} and the  central rise of the velocity dispersion
profile  observed  both from  the  ground  \citep*{peterson89} and  by
\emph{HST}  \citep{gerssen02,gerssen03}.  However, \citet{grabhorn92},
\citet{dull97},  and  \citet{baumgardt03}  have presented  dynamically
evolving   models   for   M15   that   provide  good   fits   to   the
surface-brightness and velocity-dispersion profiles \emph{without} the
inclusion  of  a  central  black  hole.   Instead,  in  these  models,
centrally concentrated populations of massive white dwarfs and neutron
stars  dominate the  gravitational  potential in  the central  region.
After  some   initial  uncertainty   caused  by  plotting   errors  in
\citet{dull97},  and  rectified  in \citet{dull03},  \citet{gerssen03}
found   that  the  \citet{dull97}   models  do,   in  fact,   fit  the
\emph{HST}-STIS velocities  reported by \citet{vandermarel02}, without
the  addition  of  a  black  hole.   Using  proper  motions  of  stars
\citet{mcnamara03} also concluded there  is little direct evidence for
an  IMBH  in  M15\@.  Using equilibrium  models  \citet{chakrabarty06}
reached a similar  conclusion.  \citet{ho03} and \citet{bash08} failed
to detect X-ray or radio  emission coincident with the cluster center.
This is not necessarily inconsistent with an IMBH, since the accretion
rate   and/or   the   accretion    efficiency   are   likely   to   be
low.  Nonetheless, M15 does  not appear  to be  a strong  candidate to
harbor an IMBH.

The   population  of   $\sim10^4$  neutron   stars  included   in  the
\citet{dull97}  models  follows  from  the adoption  of  a  reasonable
initial mass  function (IMF), an  inferred total cluster mass  of $4.9
\times 10^5~\msun$, and the  assumption that \emph{all} of the neutron
stars are  retained after formation.   As we extensively  discussed in
\citet{dull97}, this  last assumption is the least  certain, given the
likelihood that a substantial fraction  of the neutron stars formed in
clusters  will be  immediately  ejected by  their  large birth  kicks.
\citet{pfahl02}   have  examined   this  issue,   noting   that  while
consideration  of birth  kicks  suggests that  the retention  fraction
should be  in the range $1-8\%$,  X-ray evidence suggests  that it may
approach 20\%  in some clusters,  producing populations of up  to 1000
neutron  stars  in  rich  clusters.  \citet{drukier96}  found  similar
estimates for  retention fractions of neutron stars  born in binaries.
\citet{gerssen03} argued  that the neutron star  population adopted in
the \citet{dull97}  models is  implausibly large and  thus that  it is
reasonable to  effectively replace part  of the inferred  neutron star
population  with a  central IMBH  of  mass $\sim  2000~\msun$.  It  is
important to note, however, that our Fokker-Planck models only require
that  there be  a population  of nonluminous  remnants with  masses in
excess  of about  1.2~\msun; it  is not  necessary that  all  of these
objects be neutron stars.

The IMF plays a critical role in the dynamical evolution of a cluster.
This is particularly  the case for a core-collapsed  cluster, in which
extreme mass  segregation causes the massive white  dwarfs and neutron
stars  to dominate  the  central region.   The  distribution of  these
massive remnants  gives us a constraint  on the upper end  of the IMF.
Dynamical  models for  clusters provide  a  key means  of probing  the
distributions of  these extremely  faint stars, which  presently elude
optical detection.

There is  a significant  increase in the  quantity and quality  of the
data available  for detailed modeling of M15  since the \citet{dull97}
models  were developed.   Thus,  we have  developed new  Fokker-Planck
models  for M15,  to fit  these data,  that address  the issue  of the
necessity of a central black hole.

\subsection{Background}

\emph{HST} imaging  studies have provided  a wealth of  information on
globular cluster structure, stellar  content, and kinematics.  This is
particularly true  for the central  regions of high  density clusters,
which   were  largely   inaccessible  from   the  ground   before  the
\emph{HST}-era, owing to severe  crowding of stellar images.  This has
allowed  an examination  of  cluster mass  functions to  unprecedented
depth  \citet{sosin97,paresce00,pasquali04}.  In  all  three of  these
studies,  multi-mass King  models were  fit to  the  mass-function and
kinematic  data to  constrain cluster  structure parameters.   In less
crowded outer regions of clusters, \emph{HST} imaging has been used to
detect  stellar masses  near the  hydrogen burning  limit of  the main
sequence  in  NGC~6397  \citep*{piotto97,king98,richer08}  and  in  M4
\citep{richer02}. In  the central regions of  clusters, these improved
observations allow us to investigate  the details of the core collapse
process.  A predicted observational  signature of core collapse is the
presence  of a central  power-law surface  density cusp  surrounding a
small, possibly unresolved core \citep{cohn80,murphy88}.  Ground-based
observations  of  surface-brightness  profiles have  detected  central
cusps that are  interpreted as being due to core  collapse in at least
21    globular   clusters    \citep*[see   e.g.][]{trager93,lugger95}.
\emph{HST} imaging  studies of star count profiles  have confirmed the
presence  of  these cusps  in  a  number  of clusters,  including  M15
\citep{yanny94, guhathakurta96, sosin-king97}, M30 \citep{yanny94-m30,
  sosin97-m30},  and NGC~6397 \citep{sosin97}.   \citet{noyola06} have
analyzed \emph{HST} profiles for  38 clusters, finding that about half
of the sample has a central power-law cusp, with a range of slopes.

An important predicted outcome of the dynamical evolution leading to
core collapse is that substantial mass segregation will occur within
the half-mass radius of a cluster \citep{murphy88, grabhorn92, dull97,
  pasquali04}. More massive stars will lose energy to lower mass stars
and sink closer to the cluster center, as the stellar energy
distribution tends towards equipartition.  

Indications that  the stellar populations in the  central cusp regions
have been modified by  close stellar encounters provide other evidence
of the occurrence  of core collapse in clusters.   These include color
gradients  in  the  sense  of inward  bluing  \citep{djorgovski93},  a
central deficit of bright  red giants \citep{stetson94}, and centrally
concentrated populations of  both blue stragglers \citep{bailyn93} and
millisecond pulsars \citep{anderson92,phinney93}.

M15   (NGC  7078)  has   long  been   regarded  as   the  prototypical
core-collapsed  cluster.  In addition  to its  central surface-density
cusp, it has other evidence of  a high rate of stellar interactions in
its    central   region    including    eight   millisecond    pulsars
\citep{anderson92},  a blue  straggler  population \citep{demarchi94},
three   low-mass   X-ray   binaries,   AC~211,  M15   X-2,   and   X-3
\citep{hertz83,white01,heinke09},   and  at   least   one  cataclysmic
variable    located    in    the   cluster    core    \citet{shara04}.
\citet{dieball08}  and  \citet{haurberg10}  have  found  evidence  for
centrally concentrated populations of  close binary candidates in M15,
including cataclysmic variables and helium white dwarfs.

\citet{grabhorn92} and \citet{dull97} have fitted Fokker-Planck models
to the surface-brightness and  velocity-dispersion profiles of M15 and
found  that it  was best  fit  with a  global mass  function having  a
power-law index  of $x=0.9$, where  1.35 is the Salpeter  index. These
models  also containned  a sizable  population of  nonluminous objects
with masses in excess of 1~\msun.  In \citet{dull97}, the most massive
of  these  objects were  taken  to be  neutron  stars  with masses  of
1.4\msun.

\citet{sosin-king97}  and \citet{piotto97}  found evidence  for strong
mass  segregation in  M15, using  the \emph{HST}  Faint  Object Camera
(FOC) and Wide Field Planetary Camera 2 (WFPC2) imaging, respectively.
At radii of  5\arcsec, 20\arcsec, and 4.5\arcmin\ they  found that the
mass function had a power-law slope of $-2.2$, 0.2, 2.1, respectively,
for stellar masses  from 0.78 to 0.60\msun\ (where  the Salpeter slope
is  1.35).  Significant  mass segregation  has  also been  found in  a
number  of  other  clusters,  including M71  \citep{richer89},  47~Tuc
\citep{anderson96,howell01},   NGC~6397   \citep{sosin97},   and   M30
\citep{sosin97-m30,    sosin97}.    \citet{sosin-king97}    fitted   a
King-Michie  model to the  observed data  set for  M15 and  found they
could explain the  observed mass segregation, but that  they could not
accurately reproduce  the mass  function in the  central cusp  nor the
gravitational  potential  of  the  cluster.   \citet{pasquali04}  have
supplemented  the   \citet{sosin-king97}  mass-function  dataset  with
\emph{HST} NICMOS  observations at 7\arcmin, and  fitted this combined
dataset together  with several  kinematic datasets with  a King-Michie
model.    The  resulting   fit  is   more  succesful   than   that  of
\citet{sosin-king97},  most likely  due  to the  inclusion  of a  much
larger population of massive  remnants.  King-Michie models, which are
non-evolving, may not accurately represent the current dynamical state
of a  post-collapse cluster such as  M15.  These models  are likely to
substantially underestimate the degree of mass segregation that occurs
in the  core and  may thus  provide a poor  constraint on  the remnant
population.

M15 provides  us with an excellent  opportunity to examine  the IMF of
globular clusters.   In all likelihood  it has undergone  little tidal
stripping   due   to   its   position   and  orbit   in   the   Galaxy
\citep{dauphole97,vesperini97,dinescu99}.   The  fact  that M15  is  a
post-core  collapse cluster  also allows  us to  probe the  upper main
sequence of  the IMF\@.  Because  the slope of  the surface-brightness
cusp is dependent  upon the mass of the  most massive stellar remnants
we  can estimate  the number  of nonluminous  remnants present  in M15
\citep{murphy88}.  M15's  low   metallicity  (Z=0.0001)  implies  that
neutron stars were formed from progenitor stars of 6.4\msun\ and above
\citep{hurley00}. From this we can estimate the slope of the IMF above
this  mass.   The  availability  of  detailed  velocity  profiles  and
millisecond pulsar accelerations for M15  also aids in the estimate of
the    amount    of    mass    in   massive    nonluminous    remnants
\citep{phinney93,anderson92,dull97}.

\subsection{Overview}

In this paper we  present dynamically evolving Fokker-Planck models of
M15 that are  more detailed than those of  previous investigations. We
fit these models to  \emph{HST} observations of stellar mass functions
at  several  radii by  \citet{sosin-king97}  and \citet{piotto97}.  In
addition  we  fit  our  models  to the  velocity  dispersion  data  of
\citet{gebhardt94,gebhardt97,gebhardt00},            \citet{drukier98},
\citet{gerssen02},  and \citet{mcnamara03}.   In \S2  we  describe our
models. The results and  a newly determined best-fitting Fokker-Planck
model for M15 are presented in \S3.  Section 4 summarizes our findings
and discusses their implications.

\section{Dynamical Models for M15}

\subsection{Approach} 

As  in our  previous  modeling of  M15 \citep{grabhorn92,dull97},  our
basic approach  is to adopt  a global stellar mass  function, generate
evolving Fokker-Planck  models for  this mass function,  compare these
models  with  the  observed  surface-density and  velocity  dispersion
profiles at a  number of epochs, and adjust  the adopted mass function
to achieve agreement with  the observations.  Our previous modeling of
M15 adopted a power-law form for most of the range of the stellar mass
function \citep{grabhorn92,dull97}.   The direct observation  of local
stellar mass functions by \emph{HST} studies provides an important new
constraint  in the  fitting process,  that allows  the use  of  a more
complex form for the mass function.  Since \emph{HST} observations are
now  able to  probe the  stellar mass  function to  near  the hydrogen
burning        limit         of        the        main        sequence
\citep*{demarchi94,cool96,king98,richer08},       modeling       these
observations requires good resolution of the mass function using 20 or
more mass  groups; in our previous  modeling we had  used seven groups
\citep{grabhorn92,dull97}.   Our  goal  in  the present  study  is  to
constrain the  form of  the global mass  function by  fitting observed
local mass functions at several radii.

Our globular cluster models  are generated using the multi-mass direct
Fokker-Planck  method,   which  is  based  on   a  smooth,  continuous
description of  the spatial  structure of a  cluster \citep{murphy88}.
In  this  approach,  each  stellar  mass group  is  represented  by  a
statistical  distribution function in  orbital energy.   The dynamical
evolution  of  these distribution  functions  due  to  the effects  of
star-star   gravitional  scattering  are   followed  in   a  diffusive
approximation,  in  which  the  cumulative effects  of  more  frequent
distant encounters  dominate those of less  frequent close encounters.
The  gravitational  potential  due  to  the  stellar  distribution  is
determined  self-consistently,  yielding  the  time evolution  of  the
structure of  the model globular  cluster.  Using this method  we were
able to  follow the evolution of  a set of initial  models through the
first  core  collapse and  the  subsequent post-collapse  evolutionary
phase using heating  via dynamically formed binaries \citep{murphy90}.
As  reviewed by \citet{goodman93}  and \citet{murphy93},  energy input
into the  cluster core from  interactions involving close  binaries is
expected to  halt core collapse and drive  a post-collapse oscillatory
phase.   Binaries  may  be  present  in the  cluster  from  primordial
formation  and/or  may  be   formed  by  two-body  tidal  capture  and
three-body  dynamical processes. M15  likely has  few binaries  in its
core due to its being in deep collapse.  Deep collapse of a cluster is
an indication that  it has ``burned'' most of  the primordial binaries
near its center \citep{gao91}.

We begin our models with  a mass-density distribution represented by a
Plummer model ($n=5$ polytrope). All models start with the mass groups
having  identical velocity  distributions, as  expected  after violent
relaxation   \citep[\S4.7.2]{BT}.   This  velocity   equipartition  is
quickly  replaced by  energy equipartition,  causing the  most massive
stars to sink to the cluster center as the cluster evolves dynamically
\citep{inagaki85,murphy88}.  Large-scale  central density oscillations
appear soon after the  initial collapse.  These oscillations have been
observed  in conducting gas-sphere  models, Fokker-Planck  models, and
$N$-body    simulations    \citep*{sugimoto83,cohn89,makino96}.     As
discussed  in our previous  work, the  core radius  lies close  to its
maximally  expanded size  over most  of an  oscillation cycle.   For a
cluster  at a  distance  of $\sim  10~{\rm  kpc}$, such  as M15,  this
corresponds to an  angular scale of $\sim 1-2''$,  which is similar to
observational upper limits  on the core radius of  M15 from \emph{HST}
imaging \citep[e.g.][]{sosin-king97}.  We  typically evolve our models
through core collapse  and into the post-collapse phase,  to provide a
wide choice of evolutionary phases with which to fit the observations.

Recently \citet{fregeau08}  has suggested  that clusters like  M15 may
still be  in a binary  burning state rather  one of deep  collapse; we
note that this results in a  similar core radius to that achieved in a
post-collapse core bounce.

\subsection{The Stellar Mass Function \label{mass_function}} 

Our models  of M15 approximate  the present-day stellar  mass function
with a set  of 20 discrete evolved mass  groups.  This mass resolution
has allowed us  to investigate the effects of  mass segregation on the
local  mass  function.   Rather  than  explicitly  include  continuous
stellar  evolution mass  loss, which  is most  important in  the early
evolutionary phases, we use  a static \emph{present-day} mass function
for all  epochs of  our models, derived  from the IMF  by transforming
post-main-sequence   stars  to   appropriate  remnant   masses.   This
present-day evolved mass function takes into account stellar evolution
that has modified the upper IMF over the lifetime of the cluster.  For
this  study we were  mainly interested  in the  final remnant  mass of
stars so that we could determine the number and mass of the objects in
each of our remnant mass bins.  To determine the initial main sequence
to final remnant  masses for these bins we  used the stellar evolution
models described by \citet*{hurley00}.   From their fitting formula we
were able to determine the mass of the remnants produced by upper main
sequence stars of  the IMF.  Five of the  twenty mass groups represent
degenerate stars,  two represent post-main  sequence stars (horizontal
branch and red giant), and  13 mass groups represent the main sequence
stars over  the mass  range 0.8 to  0.1\msun.  Using a  metallicity of
Z=0.0001 in the \citet*{hurley00} models results in significantly more
massive white dwarfs than would solar metallicity.  Also neutron stars
are produced from progenitor stars  as low as 6.43\msun\ from this low
metallicity.


\begin{deluxetable}{ccccc}
\tablenum{1} \tablewidth{0pt} \tablecaption{Characteristics of the
Stellar Mass Functions}
\tablehead{ 
\colhead{Bin} & \colhead{Stellar}      & \colhead{Mass}   
& \colhead{Type} &  \colhead{Progenitor}\\
\colhead{}    & \colhead{Mass (\msun)} & \colhead{Fraction} 
& \colhead{}     &  \colhead{Mass (\msun)}
} 
\startdata 
1 & 1.440 & 0.0051 & neutron star & 6.43 to 22.0 \cr 
2 & 1.210 & 0.0684 & white dwarf &  4.00 to 6.43 \cr 
3 & 0.950 & 0.1339 & white dwarf & 2.00 to 4.00 \cr 
4 & 0.700 & 0.1840 & white dwarf & 1.00 to 2.00 \cr 
5 & 0.570 & 0.0638 & white dwarf & 0.82 to 1.00 \cr 
6 & 0.816 & 0.0036 & red giant & \cr 
7 & 0.796 & 0.0209 & main sequence & \cr 
8 & 0.730 & 0.0674 & main sequence & \cr 
9 & 0.619 & 0.0963 & main sequence & \cr 
10 & 0.516 & 0.0620 & main sequence & \cr 
11 & 0.442 & 0.0420 & main sequence & \cr
12 & 0.378 & 0.0326 & main sequence & \cr 
13 & 0.323 & 0.0320 & main sequence & \cr 
14 & 0.276 & 0.0315 & main sequence & \cr 
15 & 0.236 & 0.0310 & main sequence & \cr 
16 & 0.202 & 0.0306 & main sequence & \cr
17 & 0.173 & 0.0301 & main sequence & \cr 
18 & 0.148 & 0.0281 & main sequence & \cr 
19 & 0.126 & 0.0211 & main sequence & \cr 
20 & 0.108 & 0.0154 & main sequence & \cr
\enddata
\end{deluxetable}

The IMF of the cluster is given by a series of power-laws of the form,
\begin{equation}
  dN \sim \left\{ \begin{array}{ll} 
     m^{-(1+x_1)} dm, & 22 \msun > m > m_1; \\ 
     m^{-(1+x_2)} dm, & m_1 > m > m_2; \\ 
     m^{-(1+x_3)} dm, & m_2 > m > m_3;\\
     m^{-(1+x_4)} dm, & m_4 > m > 0.1\msun\\
  \end{array} \right.
\end{equation}
Here  $dN$ is  the  number of  stars  contained in  the mass  interval
$(m,m+dm)$; $x_1$, $x_2$, $x_3$,  and $x_4$ are the logarithmic slopes
of the  mass function over the indicated  mass intervals \footnote{The
  parameter $x$ is  defined as $x =  - d \ln N/d \ln  m$.}; and $m_1$,
$m_2$, $m_3$, and $m_4$ are  the stellar masses where the initial mass
function is allowed to  change power-law slope.  For progenitor masses
above the turnoff  mass, we assumed a single  power-law slope that was
varied from  $x= 0.9$  to $1.3$, where  the Salpeter mass  function is
$x=1.35$.  To estimate the number of stars in the five degenerate mass
groups,  we integrated the  IMF over  the appropriate  progenitor mass
ranges  as  indicated  in Table  1.   As  the  most massive  group  of
degenerate stars, neutron stars  of mass 1.44\msun\ will determine the
slope of  the surface-density profile  and the local mass  function in
the inner parsec  of the cluster if they dominate  the mass density in
the  cusp.  We  assume that  neutron stars  originate from  stars with
initial  masses  from  6.43  to  22  solar  masses  as  prescribed  by
\citet*{hurley00}.  The properties of each stellar group are listed in
Table~1.

Changing the upper mass limit  for neutron star progenitors has only a
small effect  on the total  number of neutron stars  produced. Because
neutron   stars   receive   a   sizeable  velocity   kick   at   birth
\citep{lyne94,hansen97}, it is likely  that a significant fraction has
been ejected  from the  cluster.  We therefore  allowed the  number of
neutron stars  to be lower than  would otherwise be  presumed from the
IMF.  For  a globular  cluster  similar  to  M15, the  central  escape
velocity is typically on the order of 50 \kms. For the pulsar velocity
distribution  of  \citet{lyne94},  fewer  than 10\%  of  the  original
neutron stars  may be present \citep{pfahl02}. This  quantity might be
higher if  one assumes  that some of  the neutron stars  originated in
binary stars \citep{drukier96,davies98,pfahl02}.  An important goal of
the present study has been to determine the minimum retention fraction
for  which  we  are able  to  fit  the  M15 observations.   For  other
degenerate  groups we  assumed  that  all stars  are  retained in  the
cluster.  The stellar  mass of each white dwarf bin  is given in Table
1, along  with the assumed progenitor  masses.  The mass  of the white
dwarf    bins    were     selected    using    a    prescription    of
\citet{girardi02,hurley00}.   Using  this  method, we  determined  the
number  of remnant  stars  present  by integrating  the  IMF over  the
various range of masses above the current turnoff mass.

Stellar mass black holes are  not included in these models for several
reasons.  As with  neutron stars black holes may  receive a birth kick
\citep{dhawan07}.  If this does occur  then very few black holes would
be retained by the cluster.  Even  if there were no birth kick most of
the black  holes produce  by a typical  mass function would  likely be
ejected.   Given the  mass  difference  compared to  the  rest of  the
cluster  stars these  stellar  mass black  holes  would undergo  rapid
dynamical  evolution,  quickly  segregate  to the  cluster  core,  and
decouple  from the  rest of  the  cluster \citep{spitzer69,watters00}.
This  central cluster  of black  holes would  then form  binaries that
would then cause them to be ejected from the system a relatively short
time \citep{kulkarni93,sigurdsson93a}.

\subsection{Model Fitting}

Our  models  can  be  described  by three  different  parameters,  the
present-day total mass of the cluster, the initial half-mass radius of
the cluster,  and the stellar  mass function parameter set.   We first
considered  an IMF similar  to that  inferred by  \citet{dull97}.  The
Fokker-Planck model  was first run  with this mass function,  then the
results were  examined to determine  if a fit  could be made,  at some
epoch, to all three local mass functions found by \citet{sosin-king97}
and  \citet{piotto97} at  5\arcsec, 20\arcsec,  and  4.5\arcmin.  From
this first model we were able to gauge what adjustments were necessary
to the  IMF, the half-mass radius,  and the cluster  mass.  We revised
the   model  until   it  satisfactorily   matched   the  mass-function
observations at the three radii at some epoch.  The models were fit to
four  types  of data:  the  mass-function  data  at the  three  radii,
star-count    profiles    for    three    luminosity    ranges,    the
velocity-dispersion  profile,   and  millisecond  pulsar  acceleration
measurements.  A  model was  deemed  acceptable  if it  satisfactorily
matched all four of these data sets.

\citet{sosin-king97}  present radial  star-count profiles  for several
magnitude ranges.   For a comparison  with their results  we converted
our stellar mass  functions to luminosity functions using  the mass to
luminosity  relation  (MLR)   of  \citet{bergbusch92}.  We  assumed  a
distance  modulus  of  15.4,  identical  to  \citet{sosin-king97}  and
\citet{durrell93}.  These MLRs allow us to make a direct comparison to
the star count data of  \citet{sosin-king97} at radii of 5\arcsec\ and
20 \arcsec.  The MLR is uncertain for low metallicities and low masses
\citep{alexander97}, so  we concentrated on fitting  the mass function
for stellar masses above 0.4\msun.

\section{Models}


\begin{figure}[t]
\epsscale{1.15}
\plotone{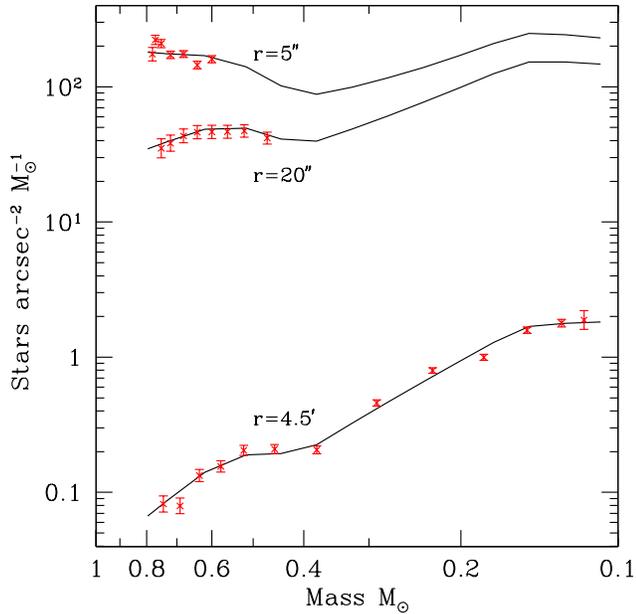}

\caption{The fit to the stellar mass function of M15.  From top to
  bottom the radius of sampling for the mass functions are 5\arcsec,
  20\arcsec, and 4.5\arcmin. The observational data are adapted from
  \citet{sosin-king97,piotto97}. The model fit at 4.5\arcmin is an
  average of three radii, 3.90\arcmin, 4.28\arcmin, and 5.46\arcmin
  corresponding to HST WFPC2 chips W2, W3, and W4, respectively.}
\label{fig1}
\end{figure}


\begin{figure}[t]
\epsscale{1.15}
\plotone{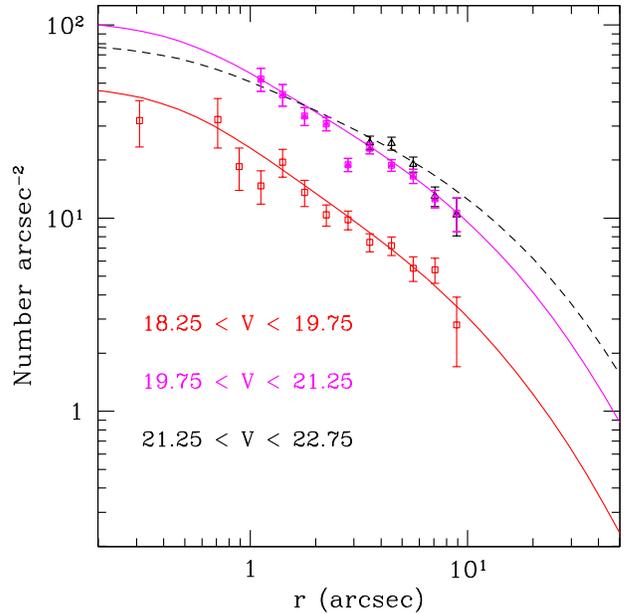}
\caption{Star count profiles for three luminosity ranges compared to
  observational data of \citet{sosin-king97}.  The lower solid curve is for
  V magnitudes between 18.25 and 19.75 ($M = 0.80\,\msun$).  The
  upper solid curve is for V magnitudes between 19.75 and
  21.25 ($M = 0.73\,\msun$).  The dashed curve is for V magnitudes
  between 21.25 and 22.75 ($M = 0.62\,\msun$).} \label{fig2}
\end{figure}

The  best-fitting model  has  a mass  of  $4.4\times10^5$\msun\ and  a
half-mass radius of 3.4 parsecs. This mass is similar to that found by
\citet{mcnamara04}. The model  was found to fit best  in the post-core
collapse phase.  While a fit of the model at the initial collapse gave
reasonable results,  we found that the post-collapse  phase provided a
better fit to  both the local mass functions  and the star-count data.
We  tried  to fit  a  number  of models  at  several  points during  a
post-collapse phase,  including both expansion  and recollapse phases.
We found that  the maximally expanded model could  not account for the
measured pulsar  accelerations.  In addition  to the finding  the best
epoch  to fit  the  model  we found  the  slope of  the  upper IMF  is
constrained by the number of  neutrons and massive white dwarfs needed
to flatten the surface-brightness profile of the cluster and produce a
large enough  pulsar acceleration.  The  IMF slope could  be steepened
but then an  unrealistic percentage of neutron stars  would have to be
retained.  The  break points in  the mass function below  the turn-off
mass  are primarly  constrained  by  the slope  of  the observed  mass
function  at 4.5\arcmin.   Break points  significantly  different than
those  listed in  Table  2 led  to a  poor  fit of  the observed  mass
function at  all radii.  Our best-fitting model,  which corresponds to
an  expanding phase,  is shown  in Figures  1---7.   The mass-function
slopes used  over the four  mass intervals, described in  section 2.2,
are given in  Table 2.  Above 0.55\msun\ we assume  a slope of $x=0.9$
with a  neutron star retention of  5\%.  The 1.4\msun\  objects in the
best-fitting model make up 0.5\% of the cluster mass.  In the range of
0.55\msun\ to  0.4\msun\ the mass function  flattens significantly, to
$x=-1.5$.   This trend  reverses  for the  lower main-sequence,  below
0.4\msun,  where $x$  increases  once again  to  0.9.  Finally,  below
0.15\msun, the mass function flattens once again to $x=-1$.


\begin{deluxetable}{cc}
\tablenum{2}
\tablewidth{0pt}
\tablecaption{M15 Initial Mass Function} 
\tablehead{
\colhead{Power-law Slope} & \colhead{Mass Interval}\\
\colhead{$x$}    & \colhead{(\msun)}} 
\startdata
0.9            & $m > 0.55 $ \cr
-1.5		& $0.55 > m > 0.40$	\cr
0.9		& $0.40 > m > 0.15$\cr
-1.0		& $0.15 > m > 0.10$	\cr
\enddata
\end{deluxetable}

Figure  1  shows  the  local  mass  function  at  radii  of  5\arcsec,
20\arcsec,  and 4.5\arcmin.  Note  that the  \citet{sosin-king97} data
only  show stellar  masses  down to  0.6  and 0.5\msun\  for radii  of
5\arcsec, 20\arcsec, respectively, due to observational limitations in
the cluster  cusp described  earlier. Our model  accurately reproduces
the  mass function  at all  three  radii.  Figure  2 shows  star-count
profile   of  the   model  compared   to  the   FOC   observations  of
\citet{sosin-king97}  for   three  magnitude  intervals.    The  three
magnitudes  intervals  correspond  to  subgiants and  stars  near  the
turnoff ($18.25 < V < 19.75$) and two main-sequence groups ($19.75 < V
< 21.25$  and $21.25 < V  < 22.75$).  The model  once again accurately
reproduces both the slope and  normalization of the profile.  Figure 3
shows  the   velocity-dispersion  measurements  of  \citet{drukier98},
\citet{gerssen02}, and \citet{mcnamara03}  compared to the model.  The
model passes through most of the error bars of the velocity points. We
do  not try  to  fit  the outer  flattening  of the  \citet{drukier98}
dataset, because  this is  likely to  be an effect  of being  near the
tidal radius.   The King-Michie model  of \citet{sosin-king97} matches
the surface  density well  but provides  only a fair  fit to  the mass
function, and fails to reproduce the velocity dispersion profile.  The
King-Michie model of \citep{pasquali04} matches the surface brightness
profile  and   the  luminosity  functions,  while   also  providing  a
reasonable  fit  to the  velocity  dispersion  profile.  As  discussed
below,  this appears  to  be a  consequence  of retaining  all of  the
neutron  stars produced from  the IMF\@.   Our model  has a  much more
realistic  retention fraction  of 5\%.   This reduced  requirement for
heavy remnants, to provide the  central potential, is a consequence of
the  higher  degree of  central  concentration  that  develops in  our
evolving Fokker-Planck models relative to a King-Michie model.  Figure
5 shows the cumulative projected mass  as a function of radius for the
total mass and  several of the mass groups.   Using observed kinematic
data  for  M15  \citet{chakrabarty06}  estimated a  mass  interior  to
0.010~pc of roughly 1000\msun. We  find somewhat less mass interior to
this  radius  in  our  model.  However  \citet{chakrabarty06}  had  an
inferred total  cluster mass nearly  3 times that of  our best-fitting
model and  of \citet{mcnamara04}.  This discrepancy could  be a result
of mass segregation and our use of evolving models and the presence of
heavy remnants.


\begin{figure}[t]
\epsscale{1.15}
\plotone{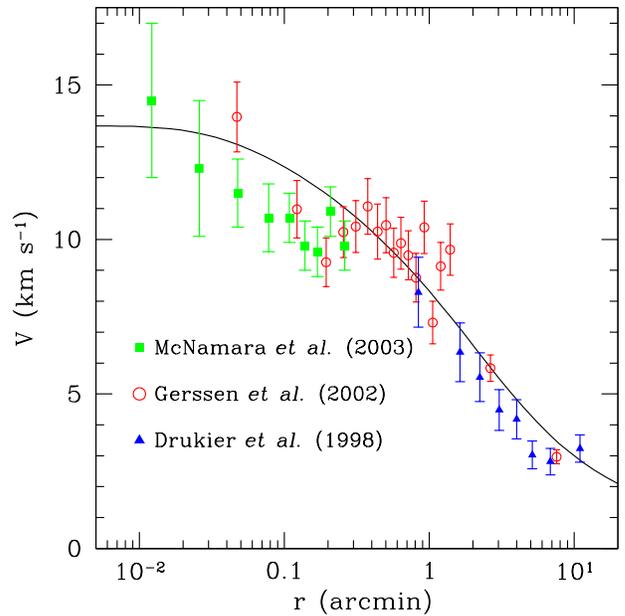}
\caption{Velocity dispersion of best-fitting model of M15 compared to
  observations of \citet{drukier98}, \citet{gerssen02}, and
  \citet{mcnamara03}.  Note that the central concentration of massive
  white dwarfs and neutron stars produces a central one-dimensional
  velocity of 13.6~\kms.}
\label{fig3}
\end{figure}

\begin{figure}[t]
\epsscale{1.15}
\plotone{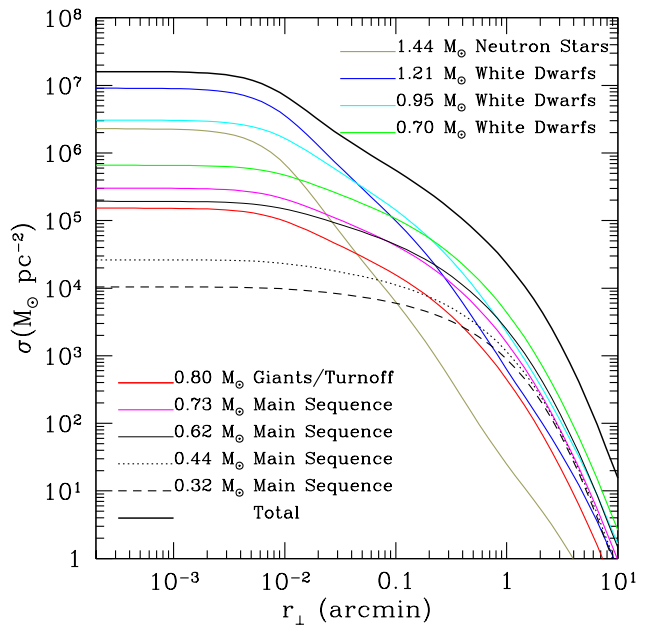}
\caption{Mass surface density profiles for the best fitting M15 model.
  Note that the region within 1\arcmin\ of the cluster center is
  dominated by white dwarfs, with massive ($M \ge 0.95\,\msun$) white
  dwarfs dominant within the inner few \arcsec.  Neutron stars are not
  dominant at any radius, due to the relatively modest retention
  fraction of 5\%.} 
\label{fig4}
\end{figure}


\begin{deluxetable}{lc}
\tablenum{3}
\tablewidth{0pt}
\tablecaption{M15 Best-Fitting Model Parameters} 
\tablehead{
\colhead{Parameter} & \colhead{Value}}
\startdata
Total Mass                                   & $4.4\times 10^5 \msun$ \cr
$r_c$                                        & $0.027$ pc \cr
$r_h$                                        & $3.4$ pc	\cr
$r_c/r_h$                                    & $0.008$ \cr
No. of Neutron Stars                         & $1560$ \cr
Projected Half-Mass Radius of Neutron Stars  & $0.17$ pc\cr
No. of Massive White Dwarfs ($m=1.21\msun$)  & $24872$ \cr
$\langle  m_* \rangle$                       & $0.43$ \msun \cr
\% Cluster Mass in Remnants                  & $45.5$\% \cr
\enddata
\end{deluxetable}

Eight    millisecond    pulsars   have    been    detected   in    M15
\citep{anderson92}. Two  of these, PSR 2127+11A and  PSR 2127+11D, are
located  1\arcsec\   from  the   cluster  center  and   have  measured
accelerations.   These accelerations  allow  us to  determine a  lower
limit  on  the  gravitational potential  \citep{phinney93,grabhorn92}.
Nonluminous remnants  dominate at  this radius, as  seen in  Figure 4.
These remnants (all  white dwarfs plus neutron stars)  make up 45\% of
the total cluster mass globally, and dominate the inner regions due to
their greater  mean mass than that  of the luminous stars.  We use the
measured  pulsar  accelarations  to  determine  the  minimum  mass  of
nonluminous remnants  required in our model. Any  dynamical model must
fit these accelerations in order  to properly map the potential in the
cusp.  Figure 6 shows  the projected gravitational acceleration versus
radius for  the best-fitting model.   A maximally expanded  model does
not  fit the  accelerations \citep{dull97}.  The  pulsar accelerations
combined  with the  fit to  the velocity-dispersion  profile  gives us
confidence that  we were  properly estimating the  nonluminous remnant
population within the half-mass  radius. \citet{dull03} found that for
high-concentration King Michie  models, the gravitational acceleration
falls  short in the  innermost regions  of the  cluster. The  model of
\citet{sosin-king97}  \citep[labeled ``King'' in  Fig.~9 of][]{dull03}
failed to fit  the accelerations (as well as  the central the velocity
dispersion  profile) because  they had  very few  massive  remnants in
their model, 21\% at 0.55\msun\ and less than 1\% in a 1.55\msun\ bin.
Because most  of their  remnants were  in a lower  mass bin,  very few
nonluminous remnants  segregated into the cusp region  of the cluster.
\citet{pasquali04} did not consider the pulsar accelerations.

Because mass segregation modifies the observed mass function, one
would be curious to see whether there is any region of the cluster
where the local mass function is a close approximation to the global
mass function.  Figure 7 shows the variation in the observed mass
function with radius.  The observed mass function changes from being
much flatter than the global mass function in the inner part of the
cluster to being steeper than the global mass function in the outer
regions of the cluster, as would be expected.  None of the curves
exactly replicate the global mass function.  However, as noted by
\citet{pasquali04}, the local mass function best matches the global
mass function near the half-mass radius.


\begin{figure}[t]
\epsscale{1.15}
\plotone{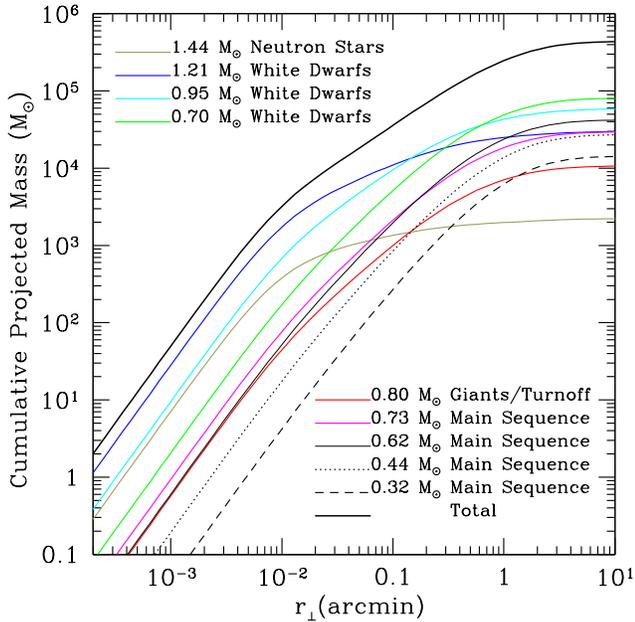}
\caption{Cumulative projected mass for the best fitting M15 model.
  The projected half-mass radii of the neutron stars and total
  cumulative mass are 0.056\arcmin\ and 0.83\arcmin,
  respectively} 
\label{fig5}
\end{figure}


\begin{figure}[t]
\epsscale{1.15}
\plotone{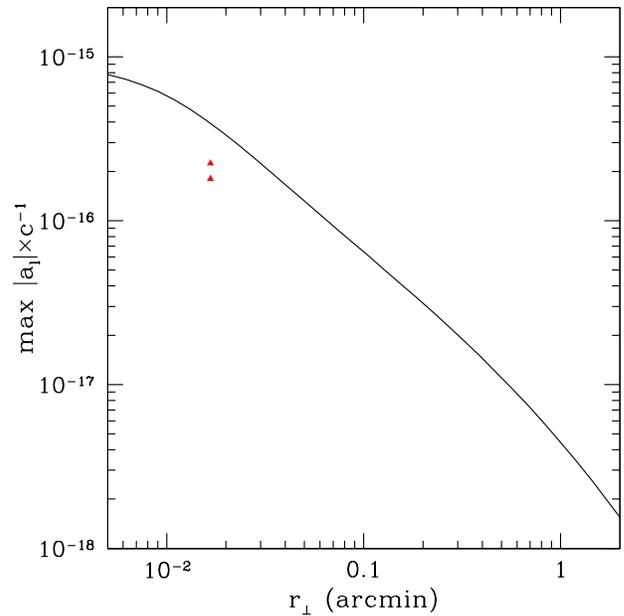}
\caption{The line-of-sight acceleration as produced by the
  mean-field gravitational potential of the best-fitting model
  compared to the observations of \citet{anderson92}. The data symbols
  indicate the minimum line-of-sight acceleration needed to account for the
  negative period derivatives of PSR 2127+11A (lower point)
  and PSR 2127+11D (upper point); the pulsars lie at nearly
  the same distance from the cluster center.}
\label{fig6}
\end{figure}

It should be pointed out that at the outermost radii of our model
cluster, i.e., radii greater than 10{\arcmin}, the local mass function
and global mass function are identical.  This is due to the very long
relaxation times at these large radii.  We do not show the mass
function here because our cluster is assumed to be isolated whereas
for a more realistic model some modification would likely have
occurred at the outermost radii due to tidal stripping by the Galaxy.

\section{Discussion}

Our inferred IMF is flatter than a Salpeter IMF over the entire
initial mass range, with a slope of $x=0.9$ for $M>0.55\,\msun$.  We
were able to fit the star count profiles, local mass functions,
kinematic data, and millisecond pulsar accelerations with a neutron
retention fraction of 5\% for neutron stars created for stars with
initial masses above 6.43\msun.  Without objects with masses of
1.2\msun\ and above, the star-count profiles would have been too steep
over the radial range of 0 to 10\arcsec.  The
islope of stellar density profile of the most luminous stars depends on
the ratio of masses of the luminous stars and the dominant mass
species \citep{cohn85}.  This is given by
\begin{equation}
   \alpha = - {{\rm d \; ln} \sigma_* \over {\rm d \; ln} r} = -1.89\left(
   {m_* \over m_{dom}}\right) + 0.65
\end{equation}
Where $\alpha$ is the logarithmic slope of the projected density
profile, $m_*$ is the mass of the luminous stars (0.81\msun), and
$m_{dom}$ is the mass of the dominant mass species (1.21\msun).  Note
that this equation implies that the slope of the cusp giants should be
$-0.63$, which is close to the observed slope
\citep{sosin-king97,noyola06}.  This slope applies to the
region where massive remnants ($M>1.2\msun$) dominate the surface
density.  This corresponds to a radius inside of 5\arcsec\ for our
models, as seen in the surface-density profiles of our best-fitting
model shown in Figure 4. Outside of this radius the surface-density
slope slowly steepens and does not have a power-law form.  

The models presented here require a total population of about 1600
neutron stars, which is a dramatic reducion relative to the $10^4$
neutron stars in the \citet{dull97} models.  This appears to be much
more realistic.  Nonetheless, it is possible that some of the 1.2 and
1.4\msun\ objects may be represented by nonluminous, tightly bound
white-dwarf binaries.  In our models 32\% of the total mass of the
cluster is in white dwarf bins with masses 0.7 and 0.95 \msun. It
could be that these objects have undergone exchange reactions with
primordial binaries composed originally of lower-mass main-sequence
stars, via 3-body or binary-binary interactions \citep{sigurdsson93}.
This would not lower the number of single white dwarfs in the core
significantly, due to their large numbers, but would add to the
1.4\msun\ bin.  These binaries could not be present in large numbers
in the form of main-sequence pairs. Such main-sequence binaries would
have an magnitude of roughly 19 and would be apparent in the star
count profiles of \citet{sosin-king97}, for the magnitude range 18.25
to 19.75, in the form of a steeper profile.  If main-sequence binaries
more massive than the turnoff did represent a sizeable fraction of the
stars in this magnitude range, they would cause the profile to be much
steeper than is observed.


\begin{figure}[t]
\epsscale{1.15}
\plotone{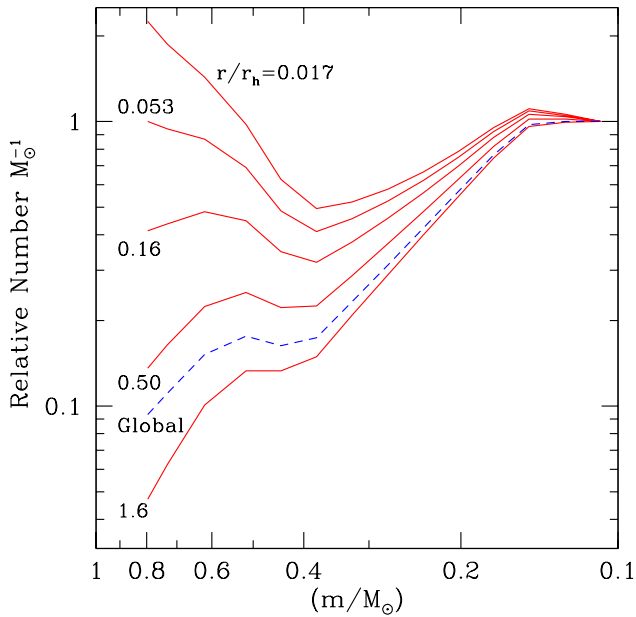} 
\caption{Radial variation of the stellar mass function as compared to
  the global mass function (dashed curve).  Note global and local mass
  functions are similar only near the half-mass radius. In this model
  $r_h=3.4$ parsec.} 
\label{fig7}
\end{figure}

\
One last explanation for the central slope of the surface density
profile of M15 is the possible presence of a massive central black
hole.  The slope of the star count profile tends to be near $-0.7$.
This is very similar to the $-0.75$ expected from a massive black hole
\citep{bahcall76}.  In the multi-mass case the slope of the projected
star-count profiles in a Keplerian potential are given by
\begin{equation}
   \alpha = - \left({ {m_* \over 4 m_{dom}} + {1\over 2}} \right)
\end{equation}
\citep*{murphy91}.  Note that if objects more massive than the turnoff
stars dominate the stellar density, then the turnoff stars will have a
slope slightly less than $-0.75$ \citep{bahcall77}.  Also
note that the minimum slope possible for any group cannot be less than
$-0.5$ if a massive black hole is responsible for the cusp. Though the
\citet{sosin-king97} data do show this trend, it should be pointed out
that in order to achieve these slopes out to 10\arcsec, the central
black hole would have to have a mass in excess of 5000\msun.  Other
evidence against a massive central black hole comes from the
velocity-dispersion measurements of \citet{gerssen02} and
\citet{mcnamara03}, which show a gradual rather than a steep rise in
the velocity dispersion.  The velocity dispersion in the Keplerian
potential of a massive black hole is expected to rise as $v \propto
r^{-0.5}$ as opposed to $v \propto r^{-0.1}$.  Thus, there is little
need to invoke the presence of an IMBH in M15 on the basis of our
model fitting.  This is in accord with the conclusion by
\citet{vandenbosch06} that the significant central dark mass
concentration in M15 can either be in the form of an IMBH, individual
objects, or a combination of these.

One useful aspect of our models is that they can help guide obsevers
as to what radii should be observed in order to best retrieve the
global mass function. We find that observations near the half-mass
radius give the best results for two reasons. First mass segregation
has not altered the mass function as much as in the inner regions and
second, tidal effects have likely had less effect at this radius than
further out.  Obtaining data at center of the cluster, 1 parsec from
the center, and at the half-mass radius appear to be the best method
for retrieving the global IMF.  These radii are nearly identical to
the data we have fit from \citet{sosin-king97}.  For clusters that are
nearer to the Sun it is important that observations be spaced
accordingly.  \citet{sosin97} shows an example of what may happen when
local mass functions are determined at two adjacent radial bins in the
cusp of a cluster.  For M30 (NGC~7099), Sosin examined local mass
functions in bins centered at 3\arcsec\ and 9\arcsec\ from the cluster
center, which corresponds to 0.1 and 0.3 parsecs, respectively. These
radii are similar to our second and third curves in Figure 7.  Figure
7 shows that the slope of the mass function above 0.4\msun\ remains
nearly constant at these small radii, even the though relative numbers
of stars have changed.  This should not be looked at as a sign of
little mass segregation. This same effect would be seen even for a
cluster in deep core collapse or bounce. This highlights the
importance of radially spacing observations in globular clusters.
\citet{sosin97} did obtain a HST FOC field at 20\arcsec, but
unfortunately the field was not usable due to the presence of a bright
giant.

Comparing our results with those of \citet{pasquali04}, we obtain a
similar slope of the IMF above 0.55\msun.  Below this mass, there are
detailed differences in the mass function, but an overall trend of
flattening or turnover.  For both sets of models, there is a
substantial amount of mass in remnants. For our models, the total mass
in remnants is 46\% of the cluster mass with 21\% of the cluster mass
in remnants above the turnoff mass.  Given the evolving nature of our
models, they are likely to provide a more realistic description of
mass segregation and the central gravitational potential of M15.

\section{Conclusions}

We have fitted dynamically evolving Fokker-Planck models to a wide
variety of data for M15, including \emph{HST} observations of the
stellar mass function at several radii, the star-count profiles, the
velocity dispersion profile, and radio measurements of millisecond
pulsar accelerations.  We infer that M15 has a relatively flat IMF
slope of $0.9$ or flatter.  With a modest retention fraction of 5\%,
we infer that M15 presently contains 1560 neutron stars.  While
of order this amount of 1.4\msun\ objects is required to fit the data,
some of these may be in the form of tightly-bound white dwarf pairs.
The central cusp region is dominated by 1.2\msun\ white dwarfs, which
represent about 7\% of the cluster mass.  For a low-metallicity
cluster like M15, a substantial number of white dwarfs are formed in
this mass range \citet{hurley00}.  Because of mass segregation, the
global mass function is not replicated at any point in the cluster.
However, the local and global mass functions come closest to each
other near the half-mass radius of the cluster.  

We conclude that there is no need to invoke the presence of an IMBH in
M15.  The presence of massive white dwarfs and neutron stars provides
the necessary central gravitational potential to fit the surface
density profile, the velocity dispersion profile, and the millisecond
pulsar accelerations.  

\acknowledgements{The authors thank Gordon Drukier for useful
  discussions and Kim Phifer for help on low-metallitcity stellar
  evolution.  BWM was partially supported by funding from the Butler
  University Institute for Research \& Scholarship.}

\singlespace

\end{document}